\documentclass{article}
\usepackage{amsmath,amssymb,url,amsthm}
\usepackage[pdftex,hiresbb]{graphicx}
\usepackage{type1cm}
\usepackage{subcaption}
\usepackage{geometry}
\usepackage{here}
\allowdisplaybreaks
\newtheorem{thm}{Theorem}[section]

\newtheorem{ass}[thm]{Assumption}

\theoremstyle{definition}

\def\d     {\displaystyle}
\def\l     {\left}
\def\r     {\right}

\def\bbE   {{\mathbb E}}
\def\bbP   {{\mathbb P}}
\def\bbR   {{\mathbb R}}

\def\tN    {\widetilde{N}}
\def\wta   {\widetilde{a}}
\def\wtb   {\widetilde{b}}
\def\wtrho {\widetilde{\rho}}
\def\wtlambda {\widetilde{\lambda}}
\def\wtS   {\widetilde{S_0}}
\def\wtsigma {\widetilde{\sigma^2_0}}
\def\bfx    {{\bf x}}
\def\ulalpha{\underline{\alpha}}
\def\ulb    {\underline{b}}
\def\whalpha{\widehat{\alpha}}
\def\Relu    {\mathop{\mathrm{Relu}}\nolimits}
\def\softplus{\mathop{\mathrm{softplus}}\nolimits}
\def\MSE    {\mathop{\mathrm{MSE}}\nolimits}
\def\RMSE   {\mathop{\mathrm{RMSE}}\nolimits}

\def\mcc   {\multicolumn{1}{c}}
\begin{document}
\title{Option pricing for Barndorff-Nielsen and Shephard model by supervised deep learning}
\author{Takuji Arai\footnote{Corresponding author} \ \footnote{
Department of Economics, Keio University, 2-15-45 Mita, Minato-ku, Tokyo, 108-8345, Japan. \\ (arai@econ.keio.ac.jp)} \\
Yuto Imai\footnote{
Faculty of International Politics and Economics, Nishogakusha University, 6-16 Sanbancho, Chiyoda-ku, Tokyo, 102-8336, Japan. \\
(y-imai@nishogakusha-u.ac.jp)}}
\maketitle

\begin{abstract}
This paper aims to develop a supervised deep-learning scheme to compute call option prices for the Barndorff-Nielsen and Shephard model
with a non-martingale asset price process having infinite active jumps.
In our deep learning scheme, teaching data is generated through the Monte Carlo method developed by Arai and Imai \cite{AI}.
Moreover, the BNS model includes many variables, which makes the deep learning accuracy worse.
Therefore, we will create another input variable using the Black-Scholes formula. As a result, the accuracy is improved dramatically. 
\end{abstract}

%
%
\section{Introduction}\setcounter{equation}{0}
We propose a supervised deep learning scheme to compute call option prices for
the non-martingale Barndorff-Nielsen and Shephard (BNS) model with infinite active jumps.
Some numerical experiments will also be conducted to confirm the accuracy of our deep learning scheme.

The BNS model is a non-Gaussian Ornstein-Uhlenbeck (OU)-type stochastic variability model that has attracted the attention of many researchers
since it was undertaken by Barndorff-Nielsen and Shephard \cite{BNS1} and \cite{BNS2}.
Now, we give a mathematical description of the BNS model.
Throughout this paper, we consider a financial market with maturity $T>0$, composed of one risky asset and one riskless asset with 0 interest rate.
The risky asset price at time $t\in[0,T]$ is described by
\begin{equation}\label{eq-BNS}
S_t:=S_0\exp\l\{\int_0^t\l(\mu-\frac{1}{2}\sigma_s^2\r)ds+\int_0^t\sigma_sdW_s+\rho H_{\lambda t}\r\},
\end{equation}
where $S_0>0$, $\rho\leq0$, $\mu\in\bbR$, $\lambda>0$, $W=\{W_t\}_{0\leq t\leq T}$ is a one dimensional standard Brownian motion, and
$H_\lambda=\{H_{\lambda t}\}_{0\leq t\leq T}$ is a subordinator without drift, that is, a driftless non-decreasing L\'evy process.
Here, $\sigma=\{\sigma_t\}_{0\leq t\leq T}$ is the volatility process defined as the square root of a solution $\sigma^2=\{\sigma^2_t\}_{0\leq t\leq T}$
to the following stochastic differential equation:
\begin{equation}\label{sigma-SDE}
d\sigma_t^2 = -\lambda\sigma_t^2dt+dH_{\lambda t}, \ \ \ \sigma_0^2>0.
\end{equation}
Now, let $N$ be the Poisson random measure of $H_\lambda$, and define its compensated Poisson random measure as $\tN(dt,dx):=N(dt,dx)-\nu(dx)dt$,
where $\nu$ is the L\'evy measure of $N$.
Then, the risky asset price process $S=\{S_t\}_{0\leq t\leq T}$ is also given as a solution to the following stochastic differential equation:
\[
dS_t = S_{t-}\l(\alpha dt+\sigma_t dW_t + \int_0^\infty(e^{\rho x}-1)\tN(dt,dx)\r),
\]
where
\[
\alpha:=\mu+\int_0^\infty(e^{\rho x}-1)\nu(dx).
\]
For more details on the BNS model, see Arai et al. \cite{AIS-BNS}, Nicolato and Venardos \cite{NV}, and Schoutens \cite{Scho}.

The Carr-Madan method, a numerical method based on the fast Fourier transform, has been commonly used to compute option prices in the BNS model.
However, it does not apply to the case where the discounted asset pricing process is not a martingale, that is, $\alpha\neq0$.
In the non-martingale case, we need to change the underlying probability measure, denoted by $\bbP$,
into an equivalent martingale measure when we compute option prices.
From the incompleteness of the BNS model, the uniqueness of equivalent martingale measures does not hold.
Hence, we need to select an appropriate one. This point will be discussed in more detail later.
The problem is that $H_\lambda$ is no longer a L\'evy process under any equivalent martingale measure.
More precisely, $H_\lambda$ has neither independent nor stationary increments.
Thus, it is impossible to describe explicitly the characteristic function of $S$ under any equivalent martingale measure.
As a result, the Carr-Madan method is unavailable for such cases.

On the other hand, conducting a Monte Carlo simulation of the BNS model with finite active jumps is not difficult, even in the non-martingale case.
Here, the BNS model is said to have finite active jumps if L\'evy measure $\nu$ is finite.
A typical example of such a case is the so-called gamma-OU type, in which $\nu$ is given by
\[
\nu(dz)=\lambda abe^{-bz}dz, \ \ \ z\in(0,\infty),
\]
where $a>0$ and $b>0$. In this case, $\nu$ is finite, and the jump parts in (\ref{eq-BNS}) and (\ref{sigma-SDE}) are given as a compound Poisson process.
Meanwhile, there had not been computational methods for the non-martingale BNS model with infinite active jumps before 
Arai and Imai \cite{AI} developed Monte Carlo simulation methods for such a case.
In particular, \cite{AI} treated the IG-OU type, a representative example with infinite active jumps in which $\nu$ is given as
\[
\nu(dz)=\frac{\lambda a}{2\sqrt{2\pi}}z^{-\frac{3}{2}}(1+b^2z)\exp\l\{-\frac{1}{2}b^2z\r\}dz, \ \ \ z\in(0,\infty),
\]
where $a>0$ and $b>0$.
Remark that the method suggested by \cite{AI} relies on the exact simulation method for $\sigma^2$
in the IG-OU case developed by Sabino and Petroni \cite{SP} and the acceptance/rejection (A/R) scheme.

As mentioned above, a measure change needs to be considered.
\cite{AI} selected the minimal martingale measure (MMM) as a representative equivalent martingale measure.
The MMM is an equivalent martingale measure appearing in local risk-minimization, which is a well-known optimal hedging strategy for incomplete markets.
Here, an equivalent martingale measure $\bbP^*$ is called the MMM if any square integrable $\bbP$-martingale orthogonal to $M$ is also a $\bbP^*$-martingale,
where $M=\{M_t\}_{0\leq t\leq T}$ is the martingale part of $S$, that is, $M$ is given as
\[
dM_t=S_{t-}\l(\sigma_t dW_t + \int_0^\infty(e^{\rho x}-1)\tN(dt,dx)\r), \ \ \ M_0=0.
\]
Under Assumption \ref{ass} below, the Radon-Nikodym density of $\bbP^*$ is given as
\begin{align*}
\frac{d\bbP^*}{d\bbP} &= \exp\bigg\{-\int_0^Tu_tdW_t-\frac{1}{2}\int_0^Tu_t^2dt+\int_0^T\int_0^\infty\log(1-\theta_{t,x})\tN(dt,dx) \\
                      &  \hspace{5mm}+\int_0^T\int_0^\infty(\log(1-\theta_{t,x})+\theta_{t,x})\nu(dx)dt\bigg\},
\end{align*}
where
\[
u_t:=\frac{\alpha\sigma_t}{\sigma_t^2+C^\rho}, \ \ \ \theta_{t,x}:=\frac{\alpha(e^{\rho x}-1)}{\sigma_t^2+C^\rho},
\]
and
\begin{equation}\label{eq-C2}
C^\rho:=\int_0^\infty(e^{\rho x}-1)^2\nu(dx)=2\rho\lambda a\l(\frac{1}{\sqrt{b^2-4\rho}}-\frac{1}{\sqrt{b^2-2\rho}}\r).
\end{equation}

\begin{ass}\label{ass}
Throughout this paper, we assume that
\[
\frac{b^2}{2}>2\l(\frac{1-e^{-\lambda T}}{\lambda}\vee|\rho|\r) \ \ \ \mbox{and} \ \ \ \frac{\alpha}{e^{-\lambda T}\sigma^2_0+C^\rho}>-1.
\]
\end{ass}

\noindent
Roughly speaking, Assumption \ref{ass} ensures that the MMM is well-defined as a probability measure.
For more details on Assumption \ref{ass}, see \cite{AI}.
Remark that the European call option at time 0 with a strike price of $K>0$ and maturity of $T>0$ is given by
\[
\bbE_{\bbP^*}[(S_T-K)^+].
\]

In this paper, we develop a supervised deep learning scheme to compute call option prices for the IG-OU type BNS model under the MMM.
To this end, we generate teaching data using the Monte Carlo method developed in \cite{AI}.
The asset price process at time 0 in the non-martingale IG-OU type BNS model includes seven variables:
$S_0$, $\alpha$, $\rho$, $\lambda$, $a$, $b$, and $\sigma_0^2$.
We need two more variables to compute the call option price: the strike price $K$ and the maturity $T$.
The training samples in deep learning will be generated by using quasi-random numbers.
In other words, the number of input data is equivalent to the dimension of the quasi-random numbers.
From this point of view, the number of variables in the BNS model is too large to develop a deep learning scheme.
Therefore, we restrict the range of each variable to around the calibrated value implemented in \cite{NV}.
Nevertheless, only restricting the range of variables does not improve the accuracy of the deep learning scheme. Therefore, we have added another idea.
Fixing the values of the seven variables in the BNS model and the maturity $T$ and regarding the option prices as a function of the strike price $K$,
we can expect that their behavior is similar to the Black-Scholes formula.
Thus, substituting $\sigma^2_0$, $K$, and $T$ into the Black-Scholes formula, we add its value to the input data.
This is a significant feature of our deep learning scheme, dramatically improving its accuracy.

Although it is possible to perform option price computation using the Monte Carlo method alone,
the trained deep learning model enables us to speed up by approximately 100 times.
It takes about 4 seconds to implement one simulation with a regular laptop.
However, the option price computation by the trained deep learning model is instantaneous.
This difference in computation time is very significant when many option prices must be computed repeatedly,
such as when performing calibration or computing volatility surface.
As a preceding study, we present Arai \cite{A}, which developed an unsupervised deep learning scheme for the BNS model.
In \cite{A}, the loss function in the neural network was defined by making use of the fact that option prices satisfy
a partial-integral differential equation.
However, the accuracy was insufficient for the non-martingale case with infinite active jumps.

The rest of this paper is organized as follows: We present the specification of our deep learning scheme in next section.
Section 3 is devoted to introducing the results of numerical experiments, and this paper concludes in Section 4.

%
%
\section{Deep learning specification}\setcounter{equation}{0}
The objective is to develop a supervised deep learning scheme for call option prices for the IG-OU type BNS model under the MMM.
Recall that a call option price at time 0 is given as a function of 9 variables: $S_0$, $\alpha$, $\rho$, $\lambda$, $a$, $b$, $\sigma_0^2$, $K$, and $T$.
We denote it by the function $F$, that is,
\[
F(S_0,\alpha,\rho,\lambda,a,b,\sigma_0^2,K,T):=\bbE_{\bbP^*}[(S_T-K)^+].
\]
In addition, the function $F_{MC}$ represents the computational result of the Monte Carlo method developed by \cite{AI}.
In this section, we introduce the neural network's structure and the results of numerical experiments.
Remark that all numerical computations in this paper are performed in MATLAB.

\subsection{Data generating}
Based on the calibration result in \cite{NV}, displayed in Table \ref{tab1}, we generate 100,000($=N$) samples first.

\begin{table}[h]
\begin{center}\caption{Calibration result in \cite{NV}}\label{tab1}
\begin{tabular}{cccccc}\hline \vspace{-3.5mm} \\
$S_0$  & $\rho$  & $\lambda$ & $a$    & $b$   & $\sigma_0^2$ \\ \hline \hline
468.40 & -4.0739 & 2.4958    & 0.0872 & 11.98 & 0.0041       \\ \hline
\end{tabular}\end{center}\end{table}

\noindent
Each value in the second column of Table \ref{tab1} is denoted by adding a tilde, e.g., $\wtrho=-$4.0739.
Generate an 8-dimensional Sobol sequence with a length of $N$, and denote it by $(\bfx_1,\dots,\bfx_N)$.
Hence, $\bfx_n$ is in $[0,1]^8$ for each $n=1,\dots,N$; we can express it as $\bfx_n=(x^1_n,\dots,x^8_n)$.
For $n=1,\dots,N$, $x^1_n,\dots,x^8_n$ are corresponding to variables $\alpha$, $\rho$, $\lambda$, $a$, $b$, $\sigma_0^2$, $K$, and $T$, respectively.
Here, we fix the value of $S_0$ to 468.40, denoted by $\wtS$.

Now, we need to set a range for each variable and transform the values of $x^1_n,\dots,x^8_n$ accordingly.
First, we set the range of $\rho$ to $\d{\l[\frac{1}{2}\wtrho,\frac{3}{2}\wtrho\r]}$, and those of $\lambda$, $a$, and $\sigma_0^2$ are also set similarly.
Hence, we transform $x^2_n$, $x^3_n$, $x^4_n$ and $x^6_n$ as follows: 
\[
\rho_n         := \frac{1}{2}\wtrho    +\wtrho x^2_n,    \ \ \ 
\lambda_n      := \frac{1}{2}\wtlambda +\wtlambda x^3_n, \ \ \ 
a_n            := \frac{1}{2}\wta      +\wta x^4_n,      \ \mbox{ and } \ 
(\sigma^2_0)_n := \frac{1}{2}\wtsigma  +\wtsigma x^6_n.
\]
Moreover, the 7th and 8th components are corresponding to $K$ and $T$, respectively.
We set then the range of $K$ to $\d{\l[\frac{1}{2}\wtS,\frac{3}{2}\wtS\r]}$, that is, we transform $x^7_n$ into $\d{K_n:=\frac{1}{2}\wtS+\wtS x^7_n}$.
On the other hand, we set the extent of $T$ from 0.01 to 1 and convert $x^8_n$ as $T_n:=0.01+0.99x^8_n$.
As for the variables $\alpha$ and $b$, we need to set the lower bounds of their ranges to meet Assumption \ref{ass}.
To this end, for $n=1,\dots,N$, we define $\ulb_n$ as
\[
\ulb_n:=1.05\times2\sqrt{\frac{1-e^{-\lambda_nT_n}}{\lambda_n}\vee|\rho_n|},
\]
which is the lower bound of $b$ in the first condition of Assumption \ref{ass} multiplied by 1.05.
We transform then $x^5_n$ so that the range of $b$ is $\l[\ulb_n,\frac{3}{2}\wtb\r]$ as follows:
\[
b_n := \ulb_n +\l(\frac{3}{2}\wtb-\ulb_n\r)x^5_n.
\]
Note that $\ulb_n<\frac{3}{2}\wtb$ holds at any time.
Next, we define $\ulalpha_n$, the lower bound of $\alpha$. To this end, we first define
\[
\whalpha_n := -\exp\{-\lambda_nT_n\}(\sigma^2_0)_n-C^\rho_n,
\]
where $C^\rho_n$ is defined by replacing $a$, $b$, $\rho$, and $\lambda$ in (\ref{eq-C2}) with $a_n$, $b_n$, $\rho_n$, and $\lambda_n$, respectively.
We define then $\ulalpha_n$ as
\[
\ulalpha_n := 0.95\whalpha_n{\bf 1}_{\{\whalpha_n\geq-3/2\}}-\frac{3}{2}{\bf 1}_{\{\whalpha_n<-3/2\}}.
\]
and transform $x^1_n$ so that the range of $\alpha$ is $\d{\l[\ulalpha_n,\frac{3}{2}\r]}$ as follows:
\[
\alpha_n := \ulalpha_n+\l(\frac{3}{2}-\ulalpha_n\r)x^1_n.
\]
Here, the absolute value of $\alpha$ is set so that it is not greater than $3/2$.

For $n=1,\dots,N$, we implement the Monte Carlo method developed by \cite{AI}
for the $n$th sample $(\wtS,\alpha_n,\rho_n,\lambda_n,a_n,b_n,(\sigma_0^2)_n,K_n,T_n)$, and denote by $MC_n$ its result, that is,
\[
MC_n := F_{MC}(\wtS,\alpha_n,\rho_n,\lambda_n,a_n,b_n,(\sigma_0^2)_n,K_n,T_n).
\]
We shall use $MC_n$ as teaching data in our scheme.
When implementing the Monte Carlo method, the number of paths and time step interval are set to 1,000 and 0.01, respectively.
Note that implementing the Monte Carlo method for 100,000 samples is very time-consuming.
It took 178 hours approximately. Therefore, it is very difficult to increase the number of samples.

Due to the large number of variables, it is not easy to construct a deep learning scheme with high accuracy.
On the other hand, we can expect that option prices for the BNS model and the Black-Scholes model exhibit similar behavior.
Therefore, we calculate the Black-Scholes formula with $(\sigma^2_0)_n$, $K_n$, and $T_n$ for $n=1,\dots,N$, and add them to the dataset.
Here, the call option price at time 0 for the Black-Scholes model with volatility $\sigma$, strike price $K$, and maturity $T$ is given as
\[
BS(\sigma,K,T) := S_0\Phi(d_+)-K\Phi(d_-)K,
\]
where the interest rate is 0, $S_0$ is the current asset price, $\Phi$ denotes the standard normal cumulative distribution function, that is,
\[
\Phi(x) := \int_{-\infty}^x\frac{1}{\sqrt{2 \pi}}e^{-\frac{y^2}{2}}dy,
\]
and we denote
\[
d_{\pm} := \frac{\log S_0-\log K}{\sigma \sqrt{T}}\pm\frac{\sigma\sqrt{T}}{2}.
\]
For $n=1,\dots,N$, we denote $BS_n := BS(\sqrt{(\sigma^2_0)_n},K_n,T_n)$.

In summary, we generate a dataset of 100,000 samples from an 8-dimensional Sobol sequence.
Each sample is composed of 10 variables: $\alpha_n$, $\rho_n$, $\lambda_n$, $a_n$, $b_n$, $(\sigma_0^2)_n$, $K_n$, $T_n$, $BS_n$ and $MC_n$.
We randomly divide our dataset into a training dataset of 98,000 samples, a validation dataset of 1,000 samples, and a test dataset of 1,000 samples.

\subsection{Neural network and learning}
For the 98,000 training samples, the first nine variables of each sample are used as input data, and the last variable, $MC_n$, will be used as teaching data.
Remark that each input data is rescaled so that the range of each variable is in the interval $[-1,1]$, e.g., $\rho_n$ is rescaled as
\[
\frac{2}{\rho_{\max}-\rho_{\min}}\l(\rho_n-\frac{\rho_{\max}+\rho_{\min}}{2}\r),
\]
where $\rho_{\max}$ and $\rho_{\min}$ are the maximum and minimum values of $\rho_n$ among all training samples.

We construct a neural network composed of six hidden layers with 200 units.
As activation functions, we place the ReLU function in the first five layers and the softplus function in the last,
where the ReLU and softplus functions are defined as $\Relu(x) := x{\bf 1}_{\{x\geq0\}}$ and $\softplus(x) := \log(1+e^x)$, respectively.

We divide the training dataset into batches of size 128 and set the number of epochs to 4,000.
That is, the number of batches is 766, and the size of the 766th batch is 80.
Adam is used as the gradient descent algorithm. The initial learning rate is 0.006, decreasing by 0.5\% every ten epochs.
Note that, for each batch, we calculate the loss value and execute the learning.
Here, we use as the loss function the half mean squared error (MSE) defined as follows:
\begin{align*}
\MSE := \frac{1}{2m}\sum^m_{i=1}(X_i-MC_i)^2,
\end{align*}
where $m$ is the bathch size, $X_i$ is the output from the neural network for the $i$th sample in the batch,
and $MC_i$ is the teaching data of the $i$th sample.
As a result, the learning is executed 3,064,000 times in total.

Using the validation dataset, we additionally calculate the MSE every 50 epochs. Here, we need to take 1,000, the size of the validation dataset, as $m$.
Since the total number of epochs is 4,000, the MSE of this type is calculated 80 times.
The output when the MSE attains its minimum will be used as the final output, which we call the trained deep learning model.
In other words, the output at the end of the 4,000 epochs is not necessarily used as the trained deep learning model.
In addition, to evaluate the whole of our scheme performance, we calculate the root mean squared error (RMSE), defined as follows, using the test dataset:
\begin{align*}
\RMSE := \sqrt{\frac{1}{m}\sum^m_i(X_i-MC_i)^2},
\end{align*}
where $m=$1,000 is the size of the test dataset, $X_i$ is the result from the trained deep learning model for the $i$th sample in the test dataset,
and $MC_i$ is the teaching data of the $i$th sample.
Constructing a neural network described above and executing the learning, we found the value of the RMSE to be 0.7578, which is sufficiently small.

%
%
\section{Numerical results}\setcounter{equation}{0}
We compute call option prices of the IG-OU type BNS model under the MMM using both the trained deep learning model
and the Monte Carlo method developed in \cite{AI} and compare the results to confirm that the accuracy of our deep learning scheme is sufficient.

To this end, we generate 7-dimensional uniform random numbers on $[0,1]$.
Each element of a random number is corresponding to the variables $\alpha$, $\rho$, $\lambda$, $a$, $b$, $\sigma_0^2$, and $T$, respectively.
We transform the values of random numbers in the same way as in subsection 2.1, and fix the value of $S_0$ at 468.40.
As for the strike price $K$, we move its value from $\d{\frac{1}{2}S_0=234.2}$ to $\d{\frac{3}{2}S_0=702.6}$ at steps of $\d{\frac{1}{100}S_0}$.

\begin{table}[htb]
\begin{center}\caption{Variable sets used in experiments}\label{tab2}
\begin{tabular}{cccccccc}\hline \vspace{-3.5mm} \\
Variable set & \mcc{$\alpha$} & \mcc{$\rho$} & \mcc{$\lambda$} & \mcc{$a$} & \mcc{$b$} & \mcc{$\sigma_0^2$} & \mcc{$T$} \\ \hline \hline
(a)          & 0.49867        & -4.71919     & 1.41919         & 0.10997   & 16.96651  & 0.00386            & 0.31475   \\ \hline
(b)          & 0.23797        & -4.69071     & 3.23799         & 0.10078   & 15.63037  & 0.00523            & 0.97049   \\ \hline
(c)          & 1.17240        & -6.72677     & 3.16246         & 0.10582   & 15.79102  & 0.00581            & 0.65818   \\ \hline
\end{tabular}\end{center}\end{table}

\begin{figure}[H]
    \begin{minipage}{0.5\hsize}\includegraphics[width=70mm]{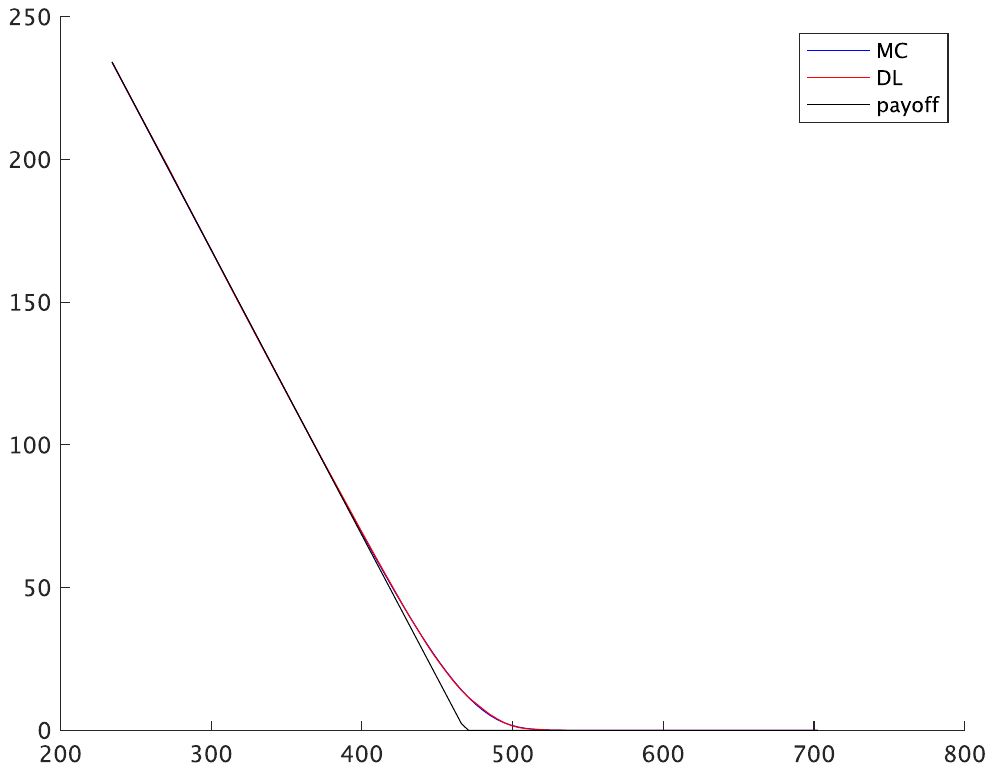}\subcaption{}\end{minipage}
    \begin{minipage}{0.5\hsize}\includegraphics[width=70mm]{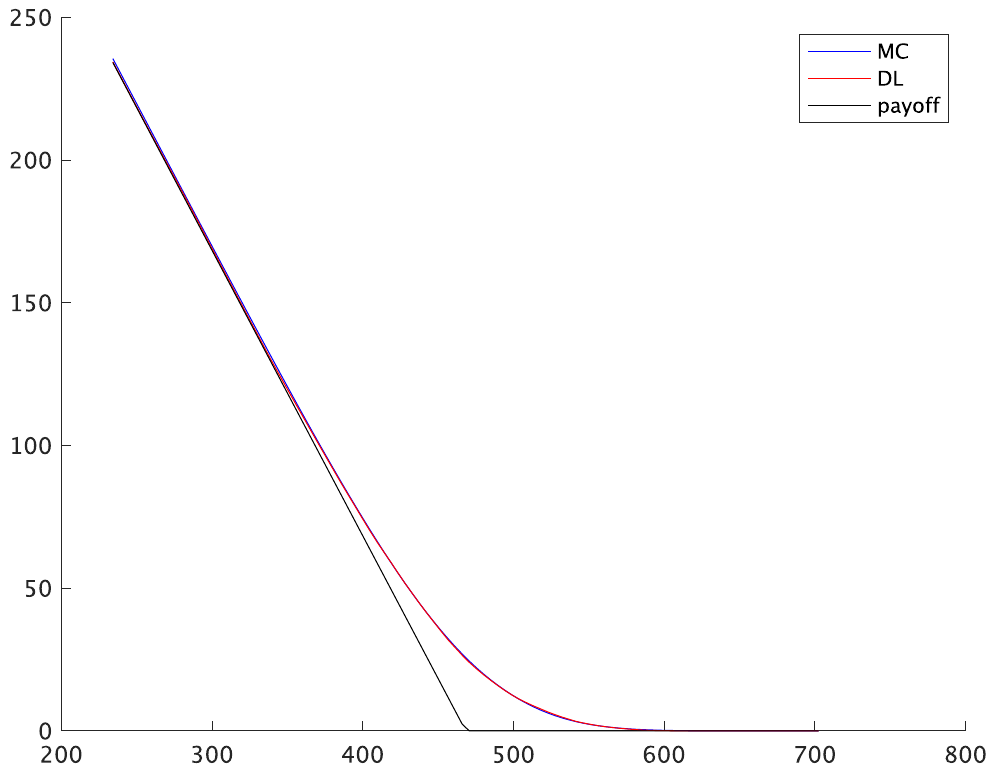}\subcaption{}\end{minipage}
    \begin{center}\begin{minipage}{0.5\hsize}\includegraphics[width=70mm]{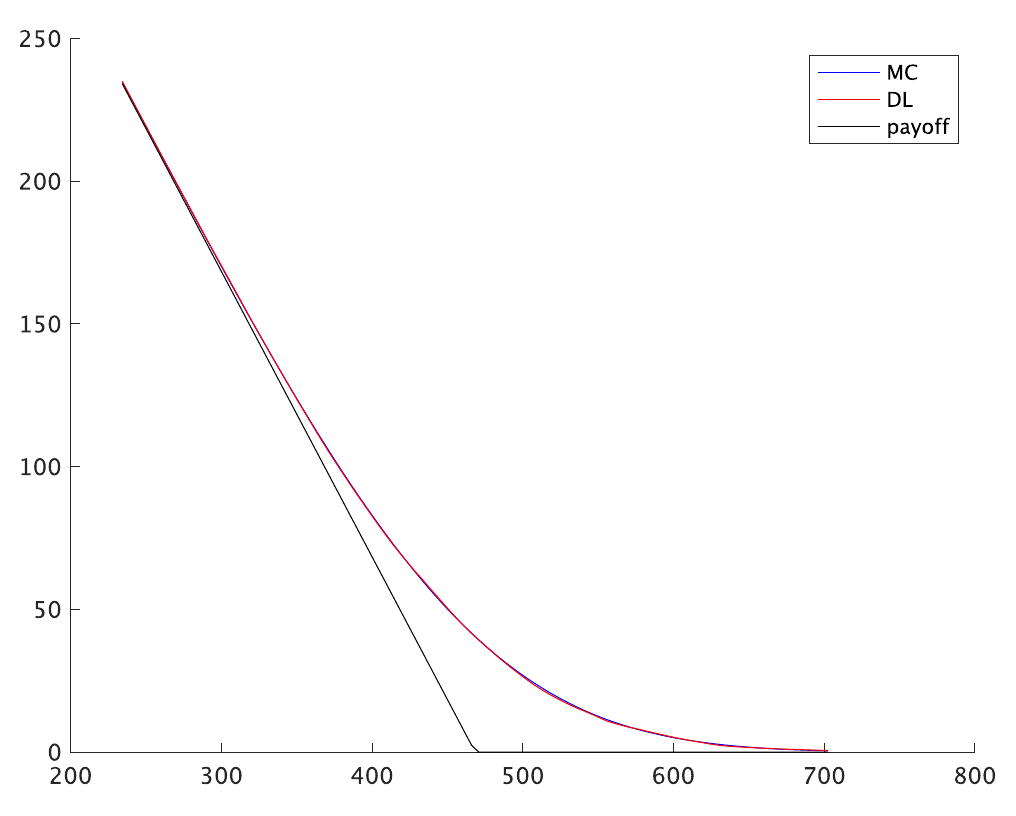}\subcaption{}\end{minipage}\end{center}
\caption{Call option prices from the Monte Carlo method (blue), the trained deep learning model (red),
and the payoff function $(S_0-K)^+$ (black) vs. strike prices.
Each panel is corresponding to the variable sets (a), (b), and (c), respectively.}\label{fig1}
\end{figure}

\begin{figure}[H]
    \begin{minipage}{0.5\hsize}\includegraphics[width=70mm]{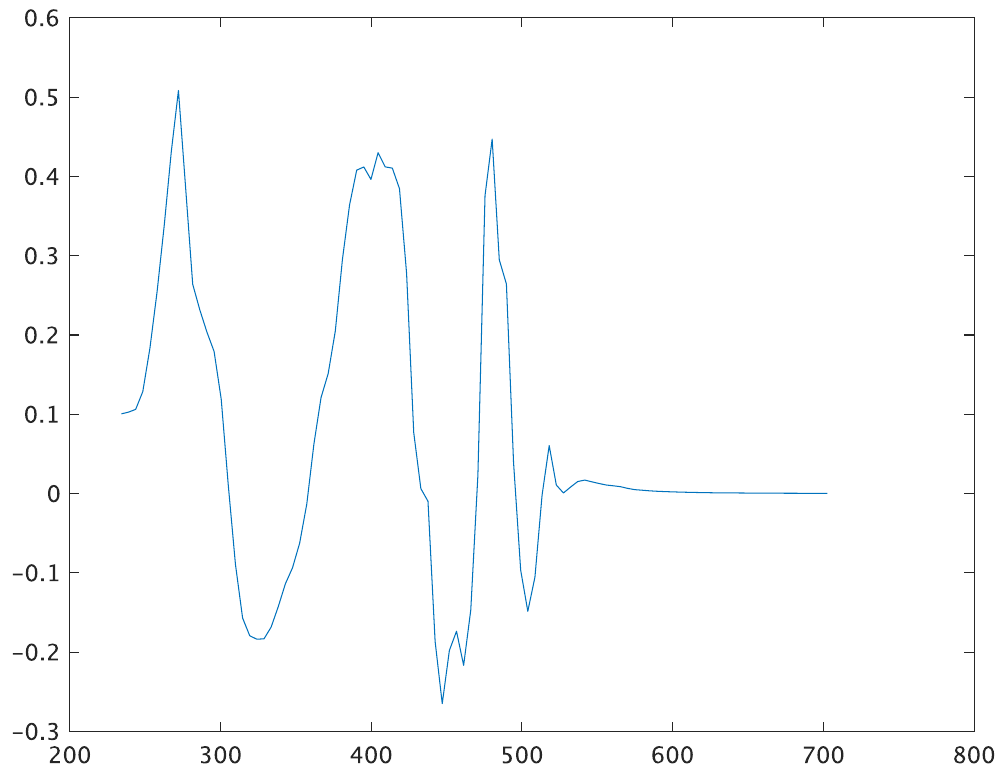}\subcaption{}\end{minipage}
    \begin{minipage}{0.5\hsize}\includegraphics[width=70mm]{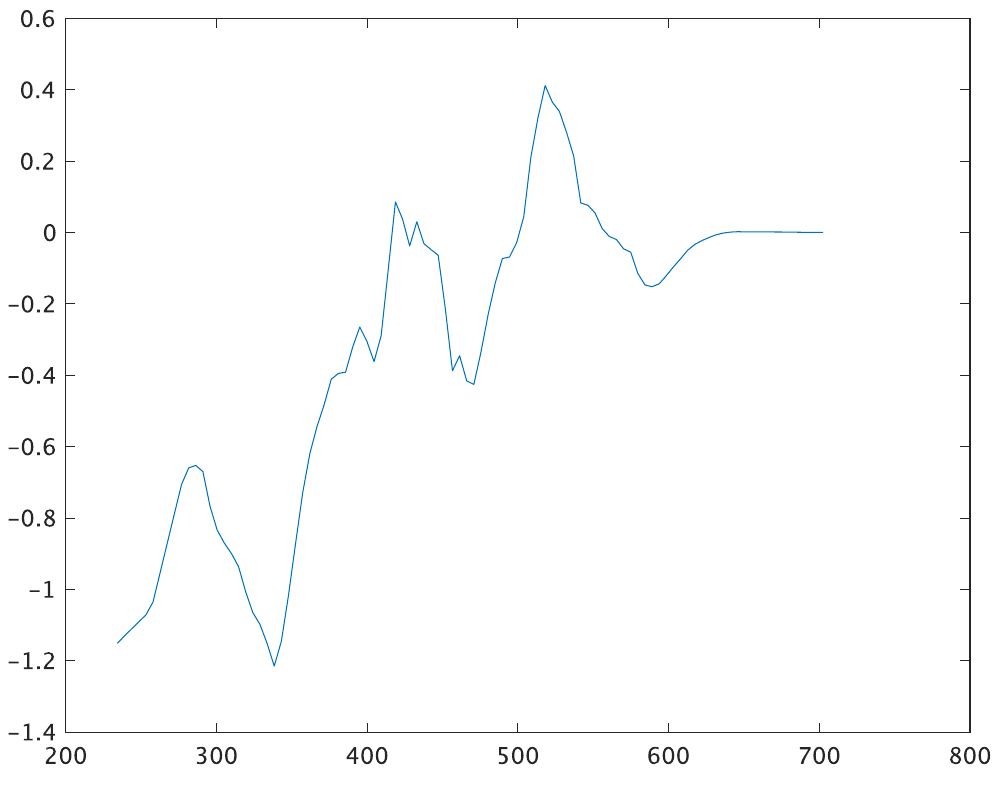}\subcaption{}\end{minipage}
    \begin{center}\begin{minipage}{0.5\hsize}\includegraphics[width=70mm]{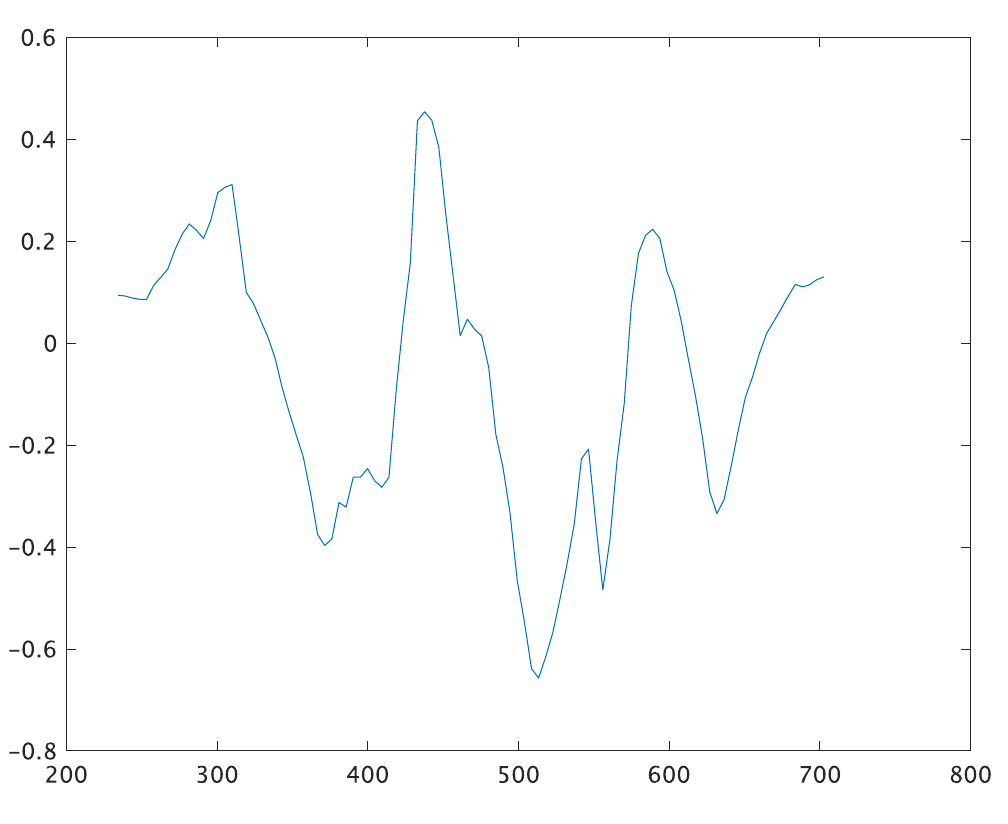}\subcaption{}\end{minipage}\end{center}
\caption{Differences between option prices from the trained deep learning model and the Monte Carlo method, vs. strike prices.}\label{fig2}
\end{figure}

\begin{figure}[H]
    \begin{minipage}{0.5\hsize}\includegraphics[width=70mm]{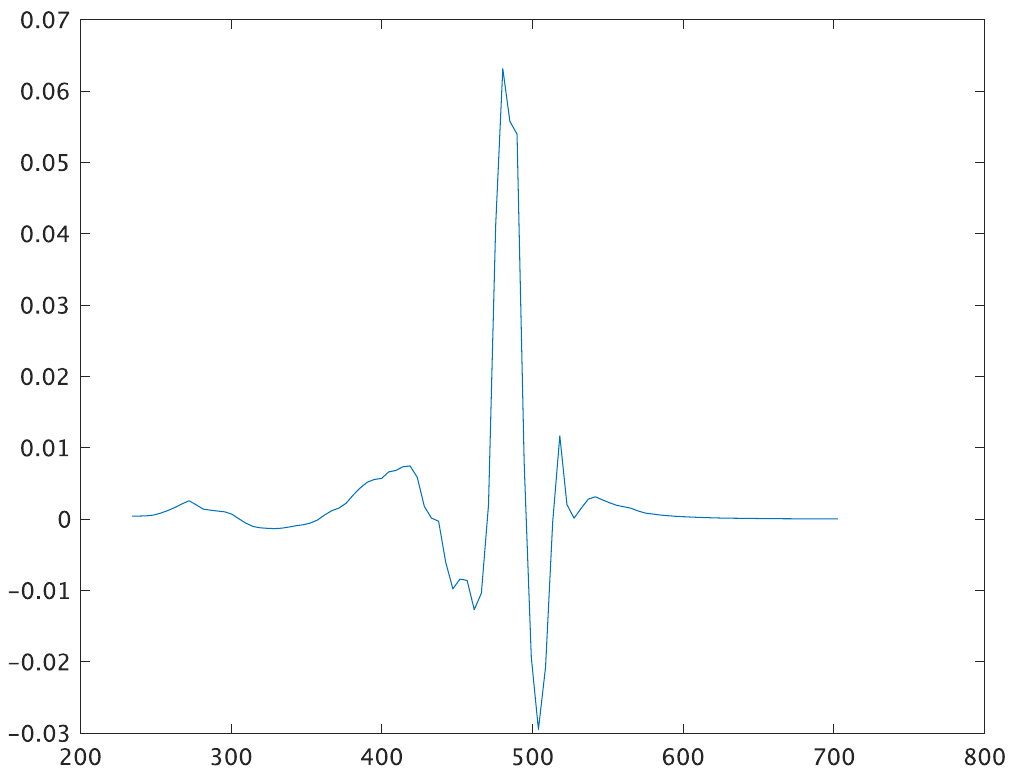}\subcaption{}\end{minipage}
    \begin{minipage}{0.5\hsize}\includegraphics[width=70mm]{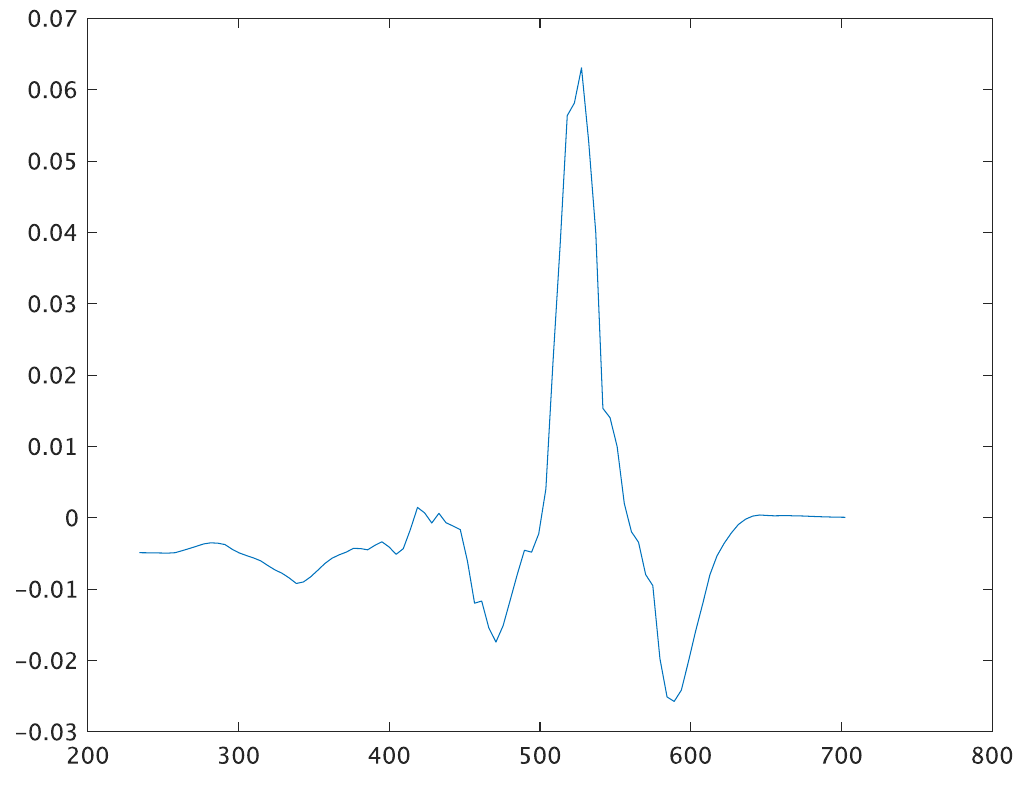}\subcaption{}\end{minipage}
    \begin{center}\begin{minipage}{0.5\hsize}\includegraphics[width=70mm]{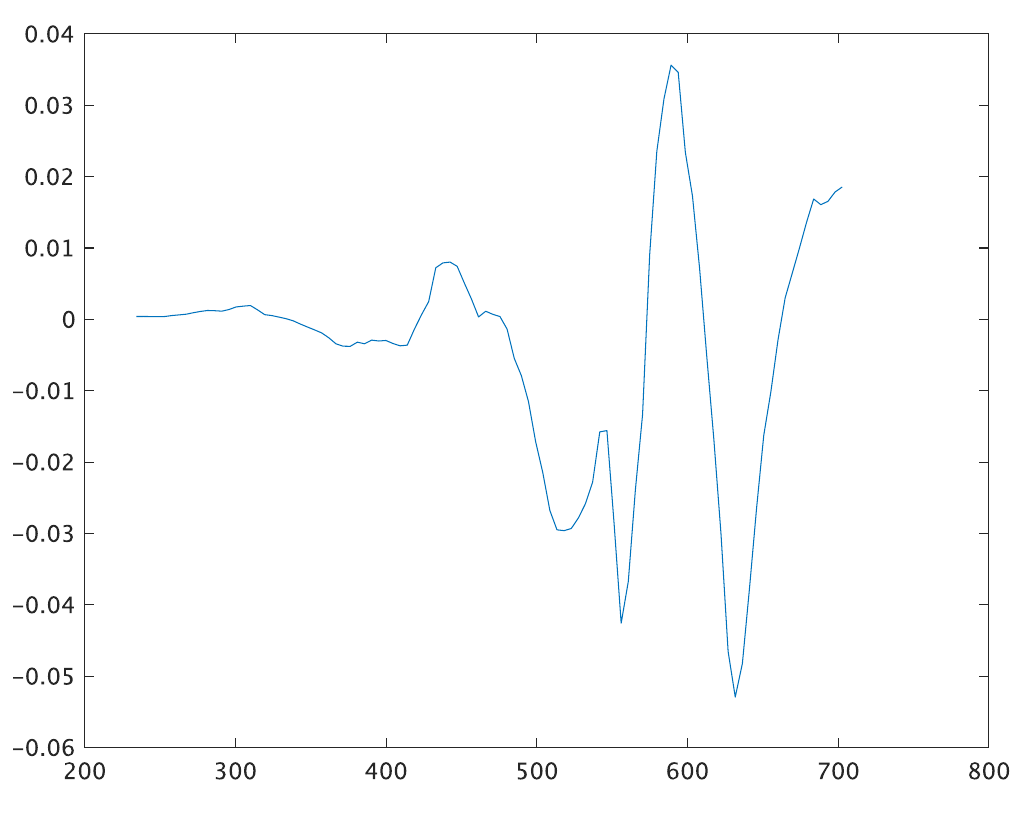}\subcaption{}\end{minipage}\end{center}
\caption{Relative errors vs. strike prices.}\label{fig3}
\end{figure}

\noindent
We introduce the results for the three sets of variables displayed in Table \ref{tab2}.
Remark that experiments for many other sets show similar results.
Figure \ref{fig1} displays the call option prices against strike prices.
Note that panels (a), (b), and (c) correspond to the variable sets (a), (b), and (c) in Table \ref{tab2}, respectively.
In each panel in Figure \ref{fig1}, the blue, red, and black curves draw option prices derived from the Monte Carlo method,
from the trained deep learning model, and the payoff function $(S_0-K)^+$, respectively.
In all panels, the blue and red curves overlap and are indistinguishable.
Figure \ref{fig2} shows the differences between the option prices using the trained deep learning model and the Monte Carlo method, that is,
we define ``difference" as
\[
\mbox{difference}:=\mbox{price by the trained deep learning model}-\mbox{price by the Monte Carlo method}.
\]
In addition, Figure \ref{fig3} displays the relative errors against strike prices,
where the relative error is defined as the difference in Figure \ref{fig2} divided by the option price derived from the Monte Carlo method, i.e.,
\[
\mbox{relative error}:=\frac{\mbox{difference}}{\mbox{price by the Monte Carlo method}\vee (K/100)}.
\]
Note that we set the lower bound of the denominator to be $K/100$ since the relative error becomes too large when the denominator is small.
In Panel (a) and (b) of Figure \ref{fig3}, the relative errors near at-the-money reach around 0.06 but are less than 0.01 otherwise.
As for Panel (c), the relative errors fluctuate in out-of-the-money but are still included in the interval $[-0.05,0.04]$.
Overall, we can say that the trained deep learning performs well.

%
%
\section{Concluding remarks}\setcounter{equation}{0}
We have developed a supervised deep learning scheme to compute call option prices for the IG-OU type BNS model under the MMM,
a representative case of the non-martingale BNS model with infinite active jumps.
In particular, we generate teaching data by using the Monte Carlo method developed by \cite{AI}.
Using the Black-Scholes formula, we created yet another input variable, which improved the accuracy of our deep learning scheme
as shown by the results of numerical experiments.

On the other hand, the generation of teaching data by the Monte Carlo method is time-consuming, so the number of samples in the dataset cannot be increased.
In addition, the BNS model has many variables, making it difficult to improve the deep learning accuracy.
Therefore, we had to restrict the range of variables in this paper.
Developing a deep learning scheme valid with broader ranges of variables is a future challenge.

\begin{center}{\bf Acknowledgments}\end{center}
Takuji Arai and Yuto Imai gratefully acknowledge the financial support of the MEXT Grant-in-Aid for
Scientific Research (C) No.22K03419 and Early-Career Scientists No.21K13327, respectively.



\begin{thebibliography}{9999}
\bibitem{A} Arai, T. (2023). Deep learning-based option pricing for Barndorff-Nielsen and Shephard model.
International Journal of Financial Engineering, 2350015.
\bibitem{AI} Arai, T. and Imai, Y. (2024). Monte Carlo simulation for Barndorff-Nielsen and Shephard model under change of measure,
Mathematics and Computers in Simulation, 218, pp.223-234.
\bibitem{AIS-BNS} Arai, T., Imai, Y. and Suzuki, R. (2017). Local risk-minimization for Barndorff-Nielsen and Shephard models,
Finance \& Stochastics, 21 , pp.551-592.
\bibitem{BNS1} Barndorff-Nielsen, O. E., \& Shephard, N. (2001).
Modelling by L\'evy processes for financial econometrics. In L\'evy processes (pp.283-318). Birkh\"auser, Boston, MA.
\bibitem{BNS2} Barndorff-Nielsen, O. E., \& Shephard, N. (2001).
Non-Gaussian Ornstein-Uhlenbeck-based models and some of their uses in financial economics.
Journal of the Royal Statistical Society: Series B (Statistical Methodology), 63(2), pp.167-241.
\bibitem{DOP} Di Nunno, G., {\O}ksendal, B., \& Proske, F. (2008).
Malliavin calculus for L\'evy processes with applications to finance. Berlin, Heidelberg: Springer.
\bibitem{G} Glasserman, P. (2004). Monte Carlo methods in financial engineering (Vol. 53, pp. xiv+-596). New York: Springer.
\bibitem{NV} Nicolato, E. and Venardos, E. (2003). Option pricing in stochastic volatility models of the Ornstein-\"Uhlenbeck type,
Mathematical Finance, 13 , pp.445-466.
\bibitem{QDZ} Qu, Y., Dassios, A., \& Zhao, H. (2021). Exact simulation of Ornstein-Uhlenbeck tempered stable processes.
Journal of Applied Probability, 58(2), pp.347-371.
\bibitem{SP} Sabino, P., \& Petroni, N. C. (2022).
Fast simulation of tempered stable Ornstein-\"Uhlenbeck processes. Computational Statistics, 37(5), pp.2517-2551.
\bibitem{Scho} Schoutens, W. (2003). L\'evy processes in finance: pricing financial derivatives, Wiley.
\end{thebibliography}
\end{document}